\documentstyle[11pt,paspconf,psfig]{article}
\input psfig.tex

\begin{document}

\title{Role of Starbursts in LINERs}
\author{J.H. Huang, Y. Chen, L. Ji , Q.S. Gu}
\affil{Astronomy Department, Nanjing University, Nanjing, 210093,China}

\begin{abstract}
We report our preliminary results on the role of starbursts in LINERs by use of
WR galaxies as a tool. The essence of this approach lies in the different
FIR-radio relation of WR galaxies from the pure AGN's.

\end{abstract}


The energetics of LINERs has been a controversial issue since their definition
by Heckman (1980). In recent years, several LINERs have been observed by HST,
or large ground-base facilities, and unambiguous evidence has been found for
their energetics (see, e.g. Ho 1998; Maoz 1998 and references therein). 
Ho favours the point of the low-luminosity end of the AGN phenomena for the
majority of LINERs, while Maoz argues for the nuclear starbursts on the other
hand (see, also Heckman, this Proceedings). We have found that the 
different FIR-radio relation of WR galaxies from the pure AGN's could serve
as a tool to classify the energetics of LINERs, a tight FIR-radio correlation
for known WR galaxies implies that both FIR and radio emission for these
sources are controlled by on-going star-formation activities.

Indeed, LINERs probably shock-excited (NGC404, NGC4569, NGC4736, NGC5055, 
NGC5194, NGC7217, Arp220, F12112+0305, F14308-1447, see, Maoz 1998, Ho 1998,
Ptak et al 1998, Larkin et al 1998, Taniguchi et al 1998) basically illustrate
a same FIR-radio correlation as WR galaxies, with lower radio emission
in general (see Fig. 1), and LINERs whose energy sources are composed of AGN
and nuclear starbursts (NGC4594, NGC6240, UGC5101, see, Ptak et al 1998,
Schulz et al 1998, Genzel et al 1997) follow the same FIR-radio correlation
as WR galaxies, with higher radio emission (see Fig. 1). Interestingly, 
LINERs powered by AGN (M87, NGC1052, NGC3998, NGC4203, NGC4278,
see Ho 1998, Ptak et al 1998, Iyomoto et al 1998, Falcke et al 1998)
show a same FIR-radio relation as the majority of
Sy1 galaxies, distinct from the FIR-radio correlation for WR galaxies (see
Fig. 2). M81 is a source that Ho claims to be
an AGN (Ho, 1998), however, Maoz thought
that the contribution from massive stars could not be excluded (Maoz, 1998).
According to our analysis,  M81 is a composite object, consistent with the 
results obtained from ROSAT data (Colbert \& Mushotzky, 1998). NGC6500 was
considered as a composite object (Barth et al 1997), though it is located 
definitely in the AGN regime in Fig 1, in accordance with the new results
by Falcke et al (1998).
Several kinds of predictions could be
given based on this method, e.g. some Sy galaxies may be powered by central
black holes and their nuclear starbursts,
as well as classification of LINERs' energetics,
and further study of the approach we proposed here
is underway.

\begin{figure}
\vspace{1.85in}
\caption{} \label{fig-1}
\end{figure}

\begin{figure}
\vspace{1.85in}
\caption{} \label{fig-2}
\end{figure}

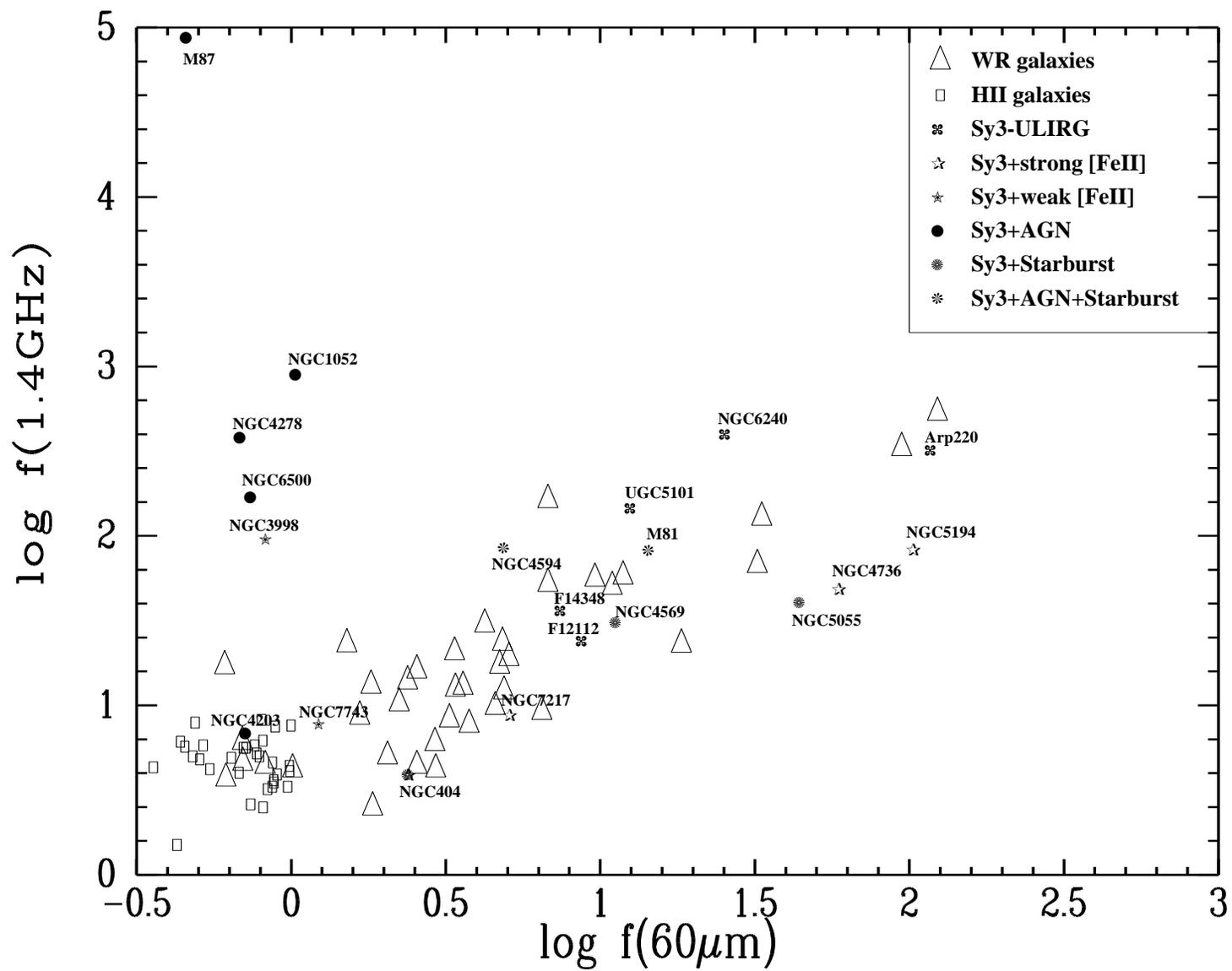
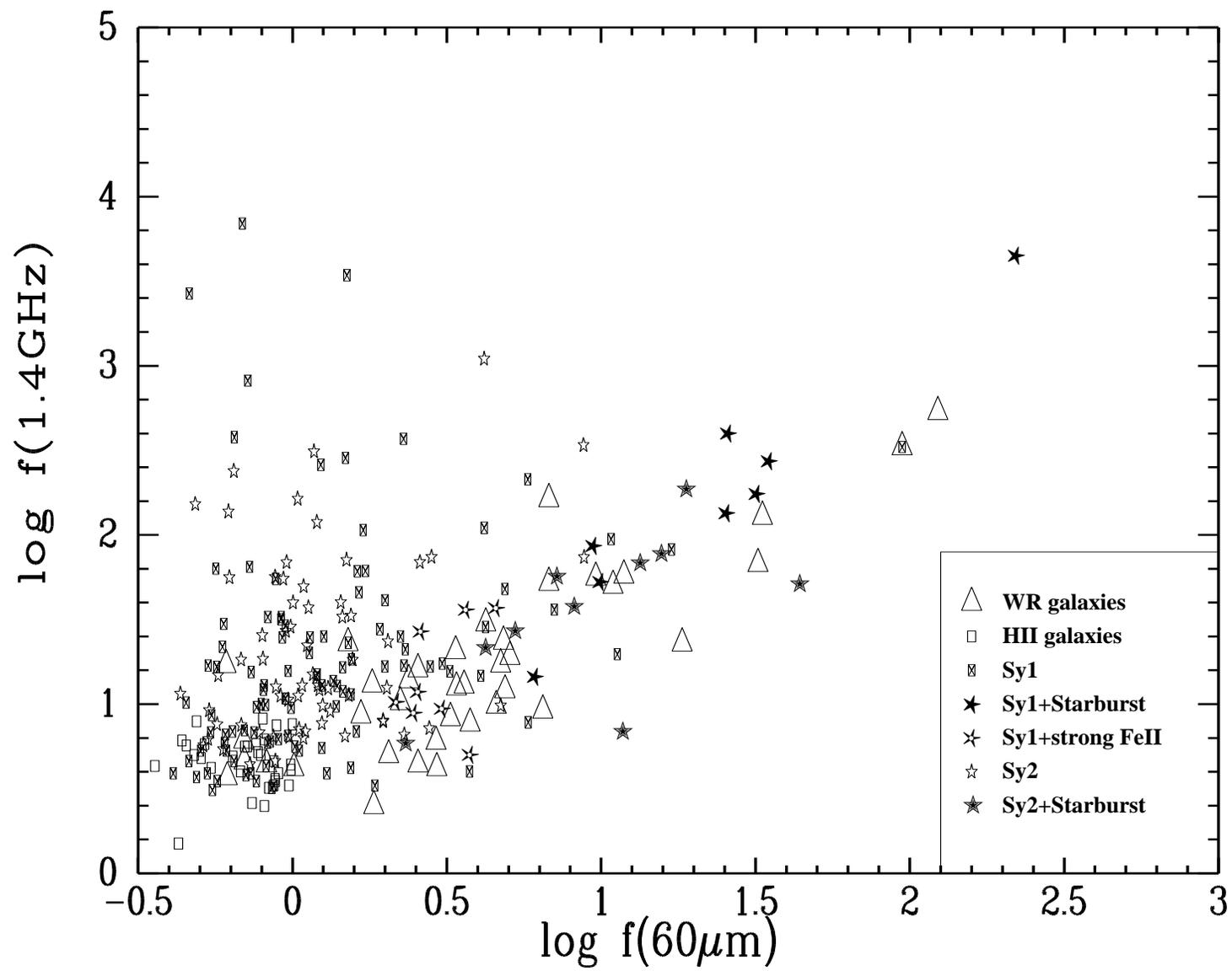
\end{document}